\documentclass[traditabstrac]{aa} 
\newcommand{\mjup}{M$_{\rm Jup}$}
\newcommand{\pmass}{M$_{\rm P}$}
\newcommand{\pap}{PA$_{\rm P}$}
\newcommand{\prad}{R$_{\rm P}$}
\newcommand{\kms}{km\,s$^{-1}$}

\usepackage{graphicx}
\usepackage{txfonts}
\usepackage{tablefootnote}
\usepackage{amsmath}
\usepackage{amssymb}

\begin{document}

  \title{Kinematics signature of a giant planet in the disk of AS 209}

   \author{
	D.\ Fedele\inst{\ref{inst_inaf}},
    F. \ Bollati\inst{\ref{inst_insubria}}
	G. Lodato\inst{\ref{inst_unimi}}
}

\institute{
INAF-Osservatorio Astrofisico di Arcetri, L.go E. Fermi 5, I-50125 Firenze, Italy\label{inst_inaf},
\and
Dipartimento di Scienza e Alta Tecnologia, Università degli Studi dell'Insubria, Via Valleggio 11, I-22100, Como, Italy\label{inst_insubria}
\and
Dipartimento di Fisica, Università degli Studi di Milano, Via Giovanni Celoria, 16, I-20133 Milano, MI, Italy\label{inst_unimi}
}

\authorrunning{Fedele et al.}

\abstract{
ALMA observations of dust in protoplanetary disks are revealing the existence of sub-structures such as rings, gaps and cavities. Such morphology are expected to be the outcome of dynamical interaction between the disk and (embedded) planets. However, other mechanisms are able to produce similar dust sub-structures. A solution is to look at the perturbation induced by the planet to the gas surface density and/or to the kinematics. In the case of the disk around AS 209, a prominent gap has been reported in the surface density of CO at $r \sim 100\,$au. A further gas gap was identified at $r \sim 200\,$au.  
Recently, \citet{Bae22} detected a localized velocity perturbation in the $^{12}$CO $J=2-1$ emission along with a clump in $^{13}$CO $J=2-1$ at nearly 200\,au, interpreted as a gaseous circumplanetary disk. We report a new analysis of ALMA archival observations of $^{12}$CO and $^{13}$CO $J=2-1$ in AS 209. A clear kinematics perturbation (kink) is detected in multiple channels and over a wide azimuth range in both dataset: the kink is visible as a displacement of the line emission in the channels map. We compared the observed perturbation with a semi-analytic model of velocity perturbations due to planet-disk interaction. Based on our analysis, the observed kink is not consistent with a planet at 200\,au as this would require a low gas disk scale height ($< 0.05$) in contradiction with previous estimate ($h/r \sim 0.118$ at $r = 100$\,au). When we fix the disk scale height to 0.118 (at $r = 100$\,au) we find instead that a planet at 100\,au induces a kinematics perturbation similar to the observed one. The kink amplitude in the various channels implies a planet's mass of 3-5\,\mjup \ . 
Thus, we conclude that a giant protoplanet orbiting at $r \sim 100\,$au is responsible of the large scale kink as well as of the perturbed dust and gas surface density previously detected. The position angle of the planet is constrained to be between 60$^{\circ}$ - 100$^{\circ}$ (East of North). 
The 200\,au gap visible in the $^{12}$CO $J=2-1$ moment 0 map is likely due to density fluctuations induced by the spiral wake. Future observations with high contrast imaging technique in the near- and mid- infrared (e.g., with JWST and/or VLT/ERIS) are needed to confirm the presence and position of such a planet.
}



   \maketitle

\section{Introduction}
In the last decade, high angular resolution observations at infrared and (sub)-millimeter wavelengths revealed the presence of sub-structures in the distribution of dust and gas in protoplanetary disks. The most common features detected by the Atacama Large Millimetre Array (ALMA) are dust and gas gaps, rings and cavities \citep[e.g.,][]{isella16, Fedele17, andrews18a, oberg21}. A common interpretation is that such substructures are the outcome of disk-planet interaction as probed e.g., by the direct detection of (partially) embedded protoplanets with infrared high-contrast imaging, as in the case of PDS 70 \citep[][]{Keppler18} and AB Aur \citep[][]{Currie22, Zhou22}. In most cases however, the direct detection remains elusive and a valuable indirect technique to infer the presence of an embedded protoplanet is to look for the perturbation in the gas surface density \citep[e.g., ][]{favre19,toci20} and/or kinematics \citep[e.g.,][]{Pinte18,teague18,Stadler23}. We report here a new analysis of ALMA archival data of the young protoplanetary system AS 209.

\smallskip
\noindent
AS 209 is a young pre-main-sequence star in the Ophiuchus star forming region at a distance of 121\,pc \citep[][]{gaiaedr3}.   
ALMA observations of AS 209 show remarkable sub-structure such as multiple dust rings and gaps \citep[e.g.,][]{Fedele18, andrews18a, zhang18} and a gas gap at $r \sim 50 - 100\,$au \citep{guzman18, favre19, alarcon21}. The highly ringed structure in the dust
continuum and the gas gap hint at the presence of a $0.2-0.3$ Jupiter mass (\mjup) planet orbiting at $r \sim 100\,$au as suggested by
hydrodynamic simulations \citep[][]{Fedele18, zhang18, favre19}. A further gas gap is detected in the CO $J=2-1$ brightness profile at $r \sim 200\,$au \citep[][]{teague18, guzman18, favre19} as well as in the near-infrared scattered light image  \citep[][]{avenhaus18}. This outer gap is detected far beyond the outermost millimeter dust continuum ring centered at $r = 120\,$au.
\citet{teague18} reconstructed the rotation curve from the fitting of the moment map and they measured a velocity deviation from keplerian motion of up to $5\,\%$ in correspondence of this outer gas gap. They ascribed such deviations to changes in the radial gas pressure gradient. A possible explanation of this finding is the presence of a giant 
planet orbiting at $r \sim 200\,$au. A valuable method to infer the presence of such a planet is the study of the gas kinematics in the CO channels map \citep[e.g.,][]{Pinte18, Pinte19}.    
Recently, \citet{Bae22} reported the detection of a kinematical perturbation in the $^{12}$CO channels map and of a circumplanetary disk in $^{13}$CO. We present a new analysis with a different methodological approach.

\begin{figure*}
    \centering
    \includegraphics[width=18cm]{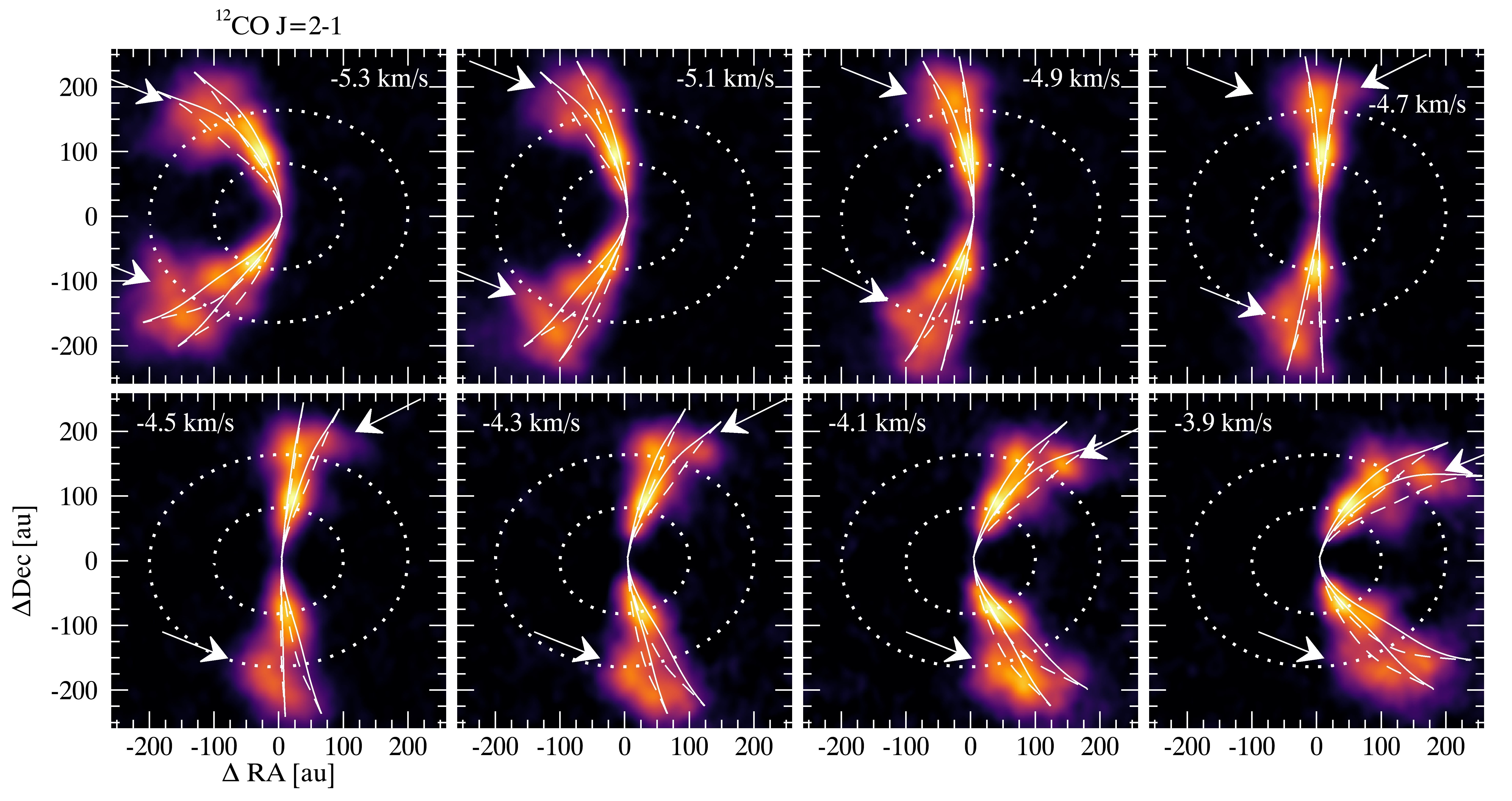}
    \includegraphics[width=18cm]{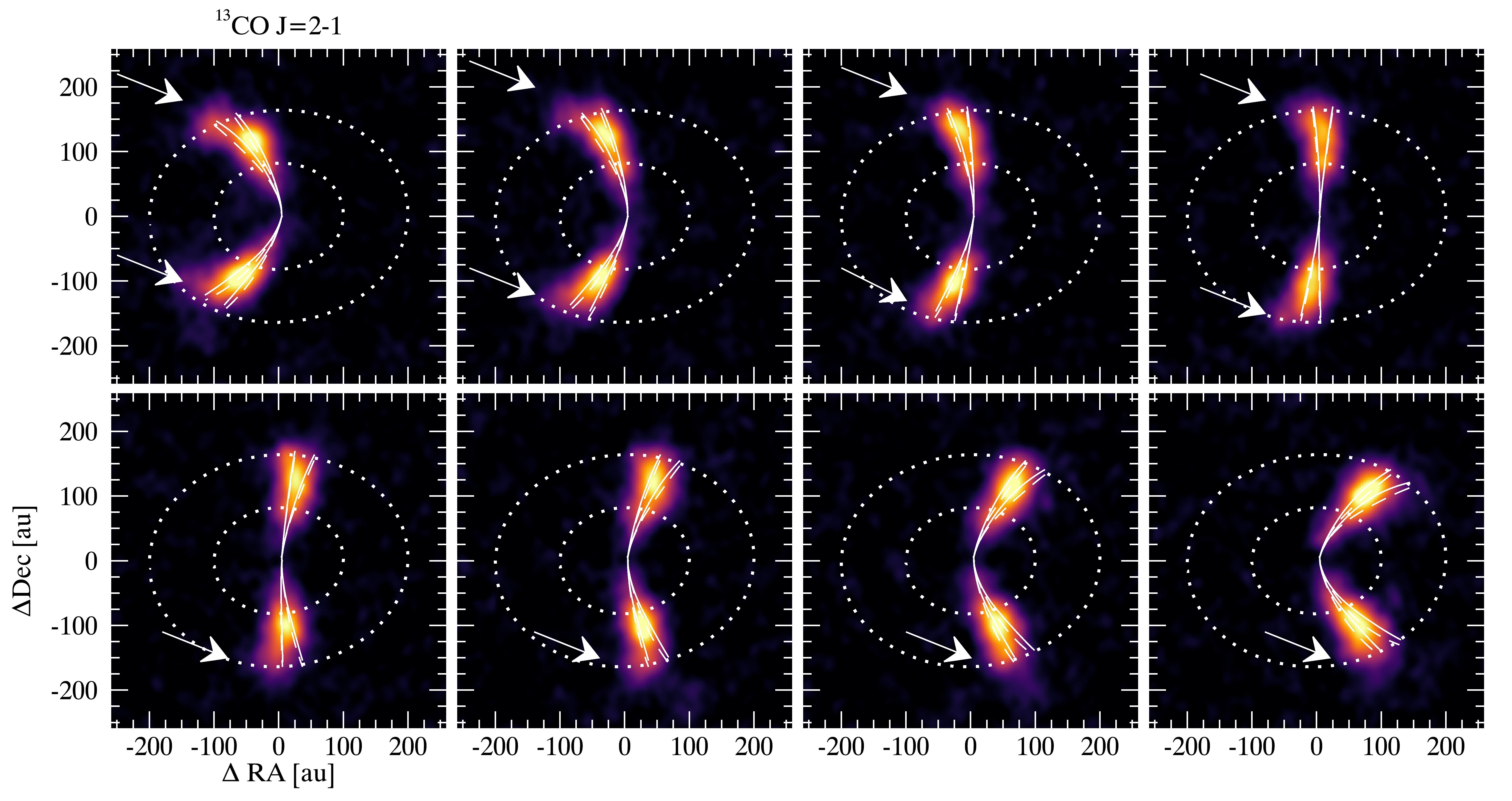}
    \caption{Selected velocity channels of $^{12}$CO (top) and $^{13}$CO (bottom) $J=2-1$ of AS 209 from \citet{oberg21, czekala21}. The dotted ellipse represents the orbital distances of $r=100$\,au and $r=200\,$au at z=0. The keplerian iso-velocity region are overlaid for the top (solid line) and bottom (dashed) surfaces (see sec.~\ref{sec:analysis}). The arrows point to the deviation from keplerian velocity. The deviation is detected in multiple channels and over a wide range of azimuth angles both in the $^{12}$CO and in the $^{13}$CO channels maps.} 
    \label{fig:data}
\end{figure*}

\begin{figure*}
    \centering
\includegraphics[width=9cm]{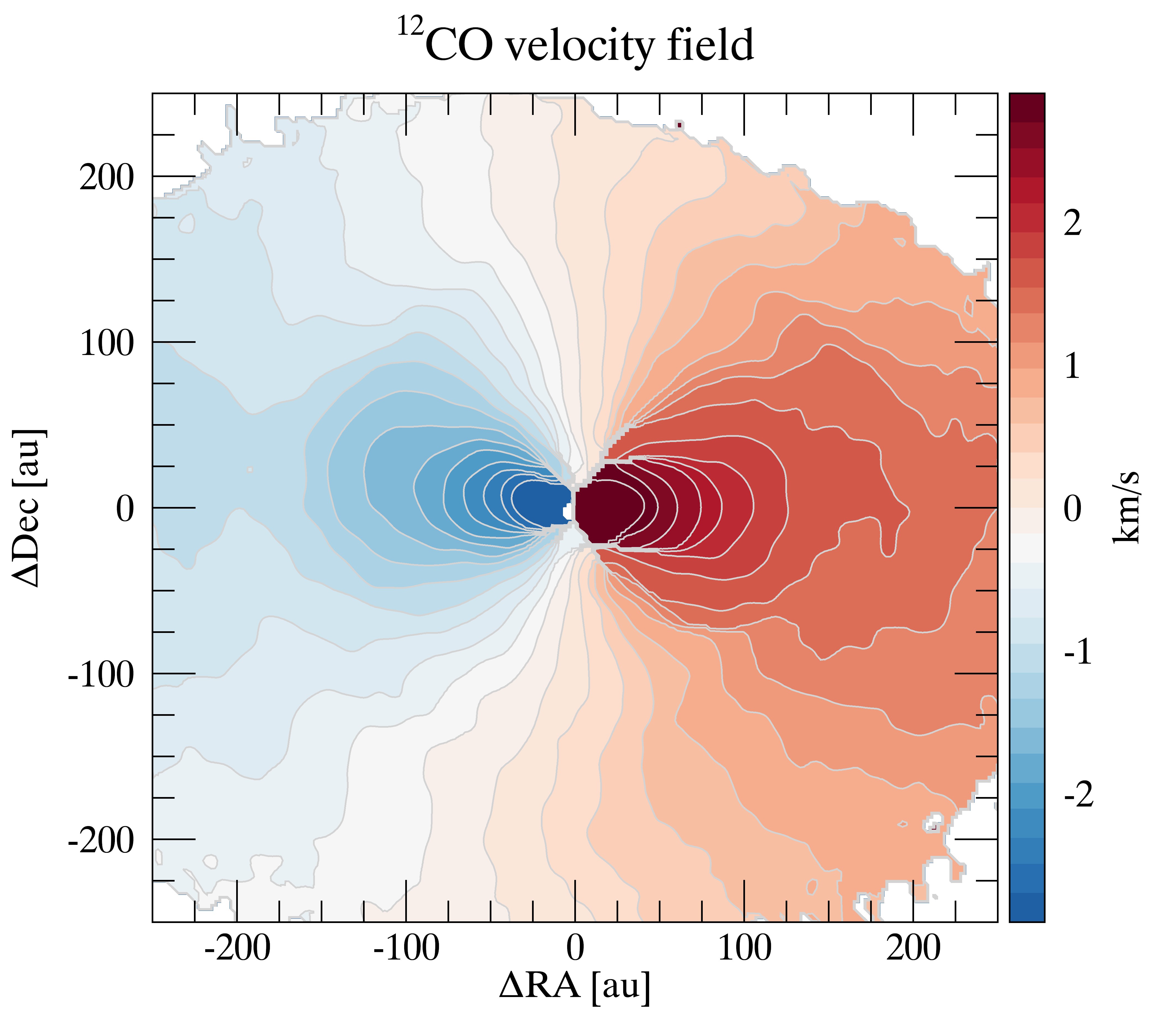}
\includegraphics[width=9cm]{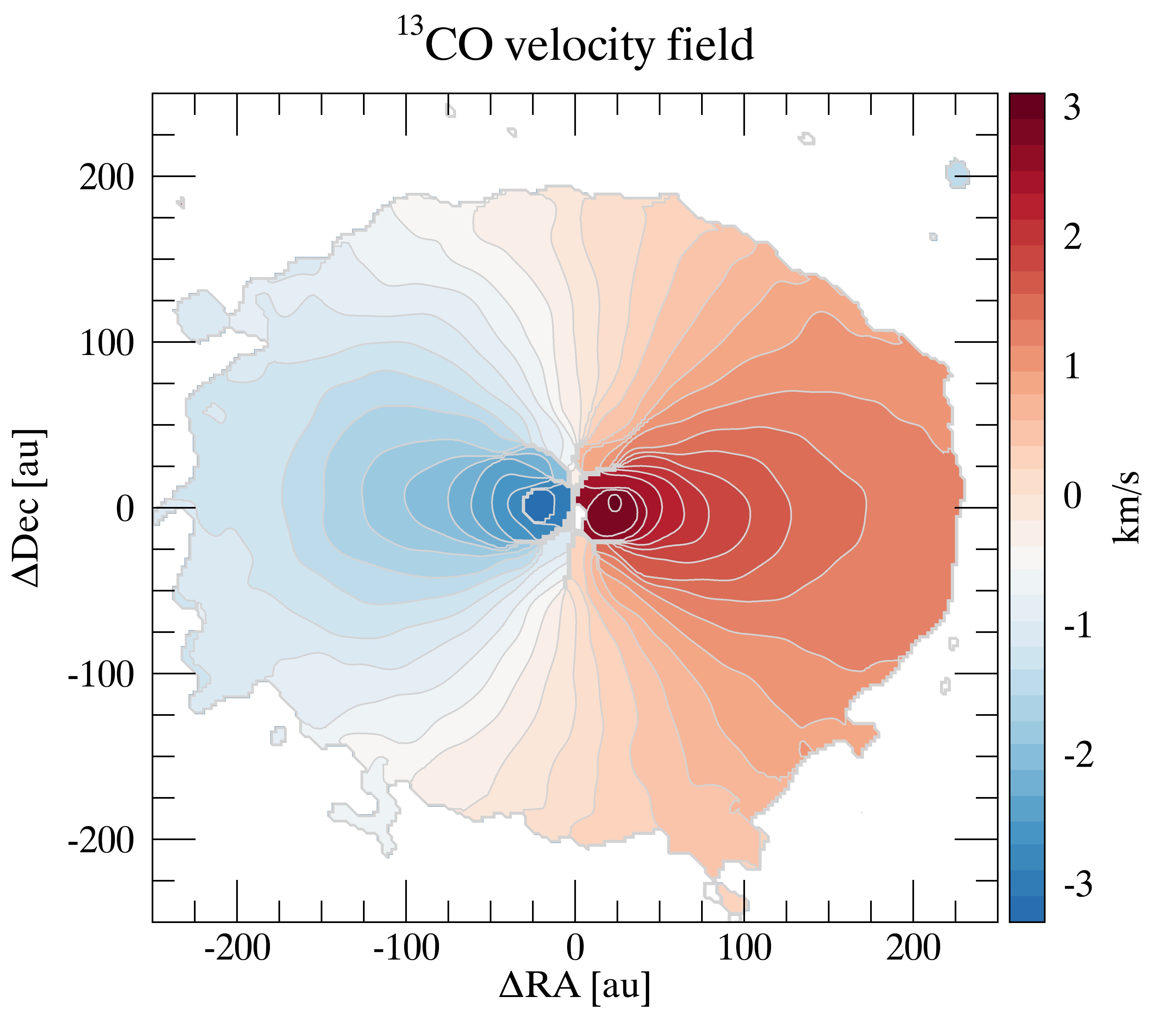}
\includegraphics[width=9cm]{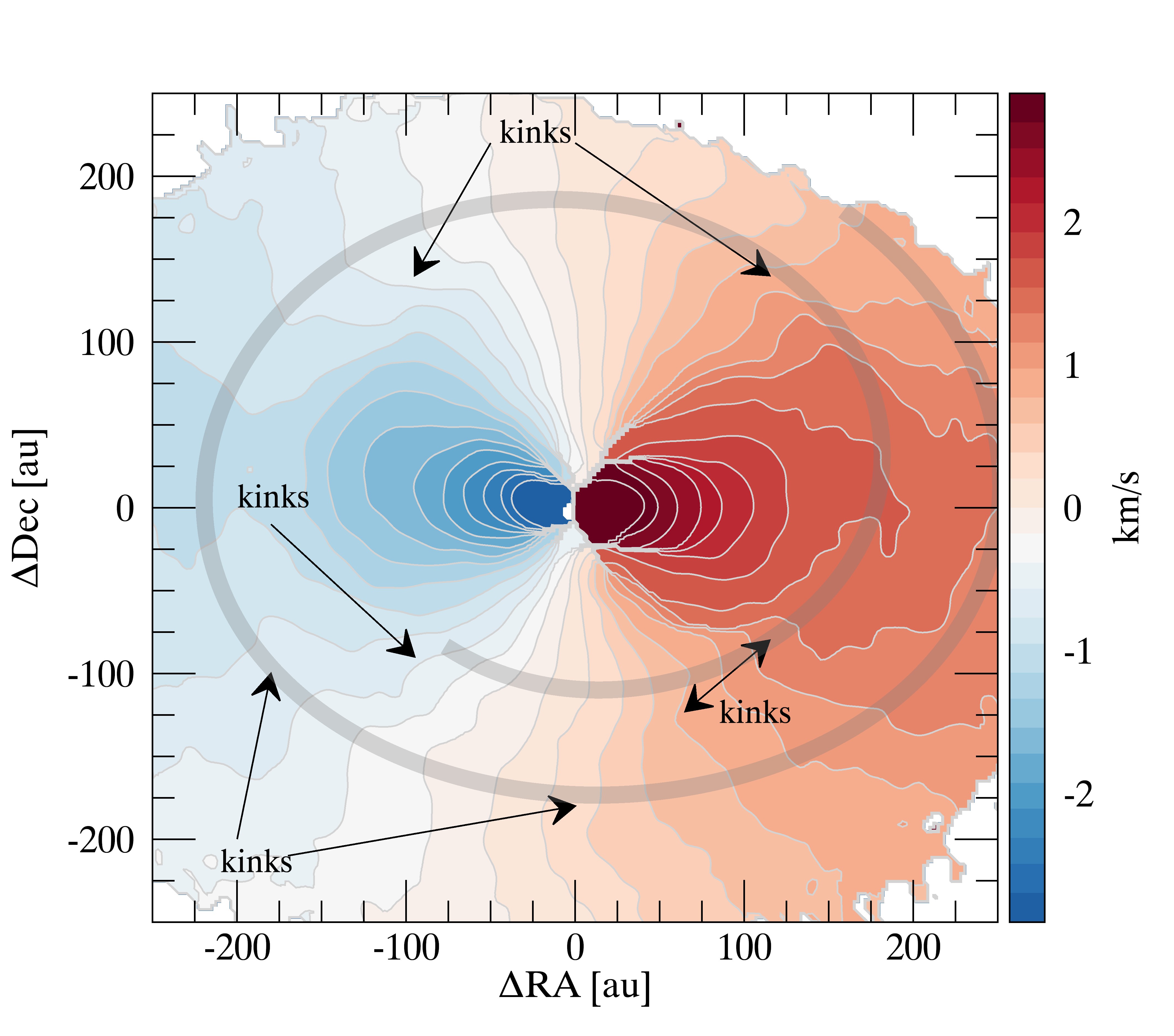}
\includegraphics[width=9cm]{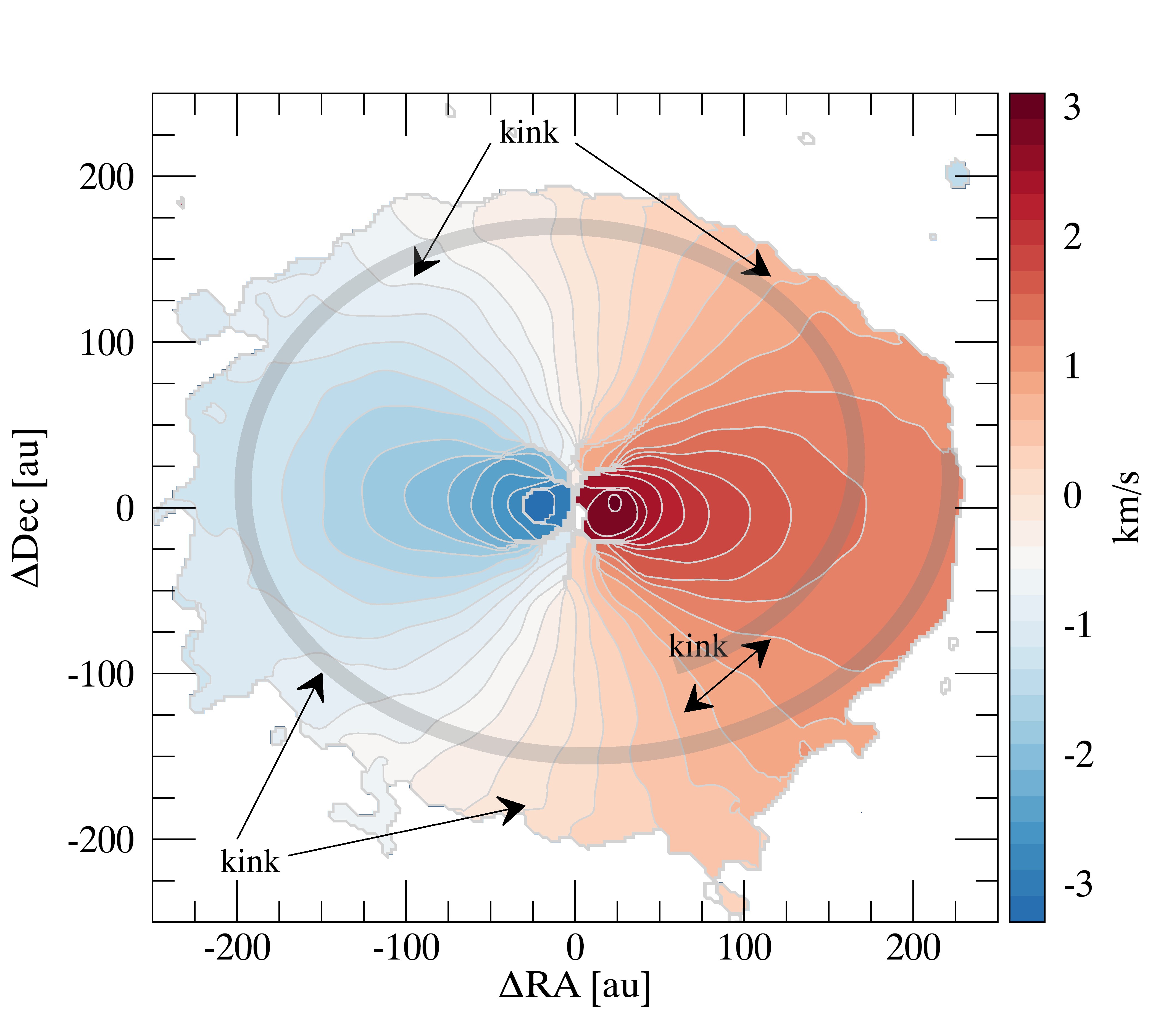}
    \caption{Velocity map of the peak intensity of $^{12}$CO (left) and $^{13}$CO (right) $J=-2-1$ in the LSR frame of the system. The bottom panels show the same maps with the most prominent kinks indicated by the arrows. The large scale structure of the kink resembles a spiral wake, tentatively represented by the thick grey line. }
    \label{fig:data_mom9}
\end{figure*}

\section{Data}
The analysis presented here is based on archival ALMA observations of $^{12}$CO and $^{13}$CO $J =2-1$ previously published in \citet{favre19} and \citet{oberg21} (MAPS ALMA program). The channel maps are consistent between the two dataset, as the MAPS data have higher signal-to-noise ratio and slightly higher angular resolution, we adopted the MAPS spectral cube for the following analysis. 
The MAPS observations and data reduction are fully described in \citet{czekala21}. For our analysis we downloaded the self-calibrated and JvM corrected dataset (following the calibration described in \citealt{Jorsater95}) with an angular resolution of 0\farcs2 and channel width of 0.2\,\kms. \citet{Casassus22} warned about the JvM correction as it can overestimate the peak signal-to-noise ratio in the restored images. For this reason we also show the same dataset without the JvM correction in appendix.
The channels map and the velocity map are shown in Figures~\ref{fig:data} and \ref{fig:data_mom9}, respectively. The latter corresponds to the velocity map of the peak intensity computed with a quadratic fit using the code \textsc{bettermoments} \citep{Teague18b}.

\section{Analysis}\label{sec:analysis}
\subsection{Emission surfaces and keplerian isovelocity curves}
We used the code "CO\_layers"\footnote{Avaliable at
https://github.com/richteague/disksurf} \citep{Pinte18} to determine the keplerian isovelocity regions of the $^{12}$CO and $^{13}$CO $J=2-1$ emitting surfaces. 
\citet{teague18} estimated the emitting surfaces adopting a power-law dependence of z (the height of the $^{12}$CO emitting layer over the midplane) on r (radial distance from the star):

\begin{equation}
    z(r) = r \cdot tan \phi   
\end{equation}\label{eq:t18}

\citet{law21} performed a different fit using a modified functional form that includes an exponential cut-off:

\begin{equation}
    z(r) = z_0 \cdot \Bigg(\frac{r}{1''}\Bigg)^{\phi} \cdot \exp\Bigg[-\Bigg(\frac{r}{r_{\rm taper}}\Bigg)^{\psi}\Bigg] 
\end{equation}\label{eq:l21}

\noindent

We adopted the fit results by \citet{law21}. The keplerian isovelocity curves are overlaid on the channel maps in Figure~\ref{fig:data} for the top (solid line) and bottom (dashed) surfaces.

The $^{12}$CO gap at $r \sim 200\,$au previously identified is clearly visible in the high velocity channels. 
A deviation from keplerian velocity ($\Delta v$) is visible in the outer disk in multiple channels and at different azimuth angles. Deviations from keplerian velocity (indicated by the arrows in Figure~\ref{fig:data}) are clearly detected in both the $^{12}$CO and $^{13}$CO dataset.
The velocity deviation are clearly visible also in the velocity map of the 2 transitions (Figure~\ref{fig:data_mom9}).

\begin{figure}[!t]
    \centering
    \includegraphics[width=8cm]{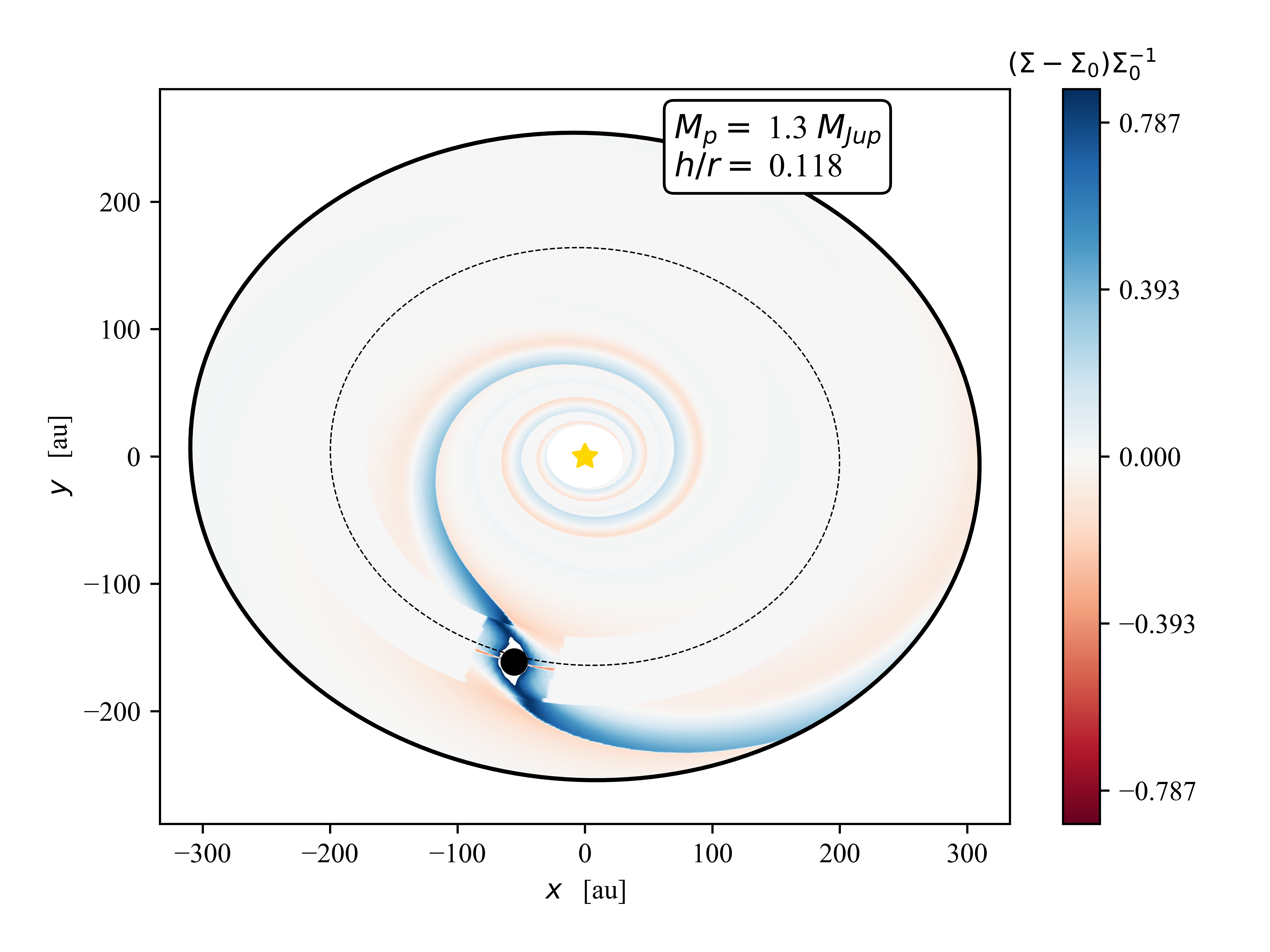}
    \includegraphics[width=8cm]{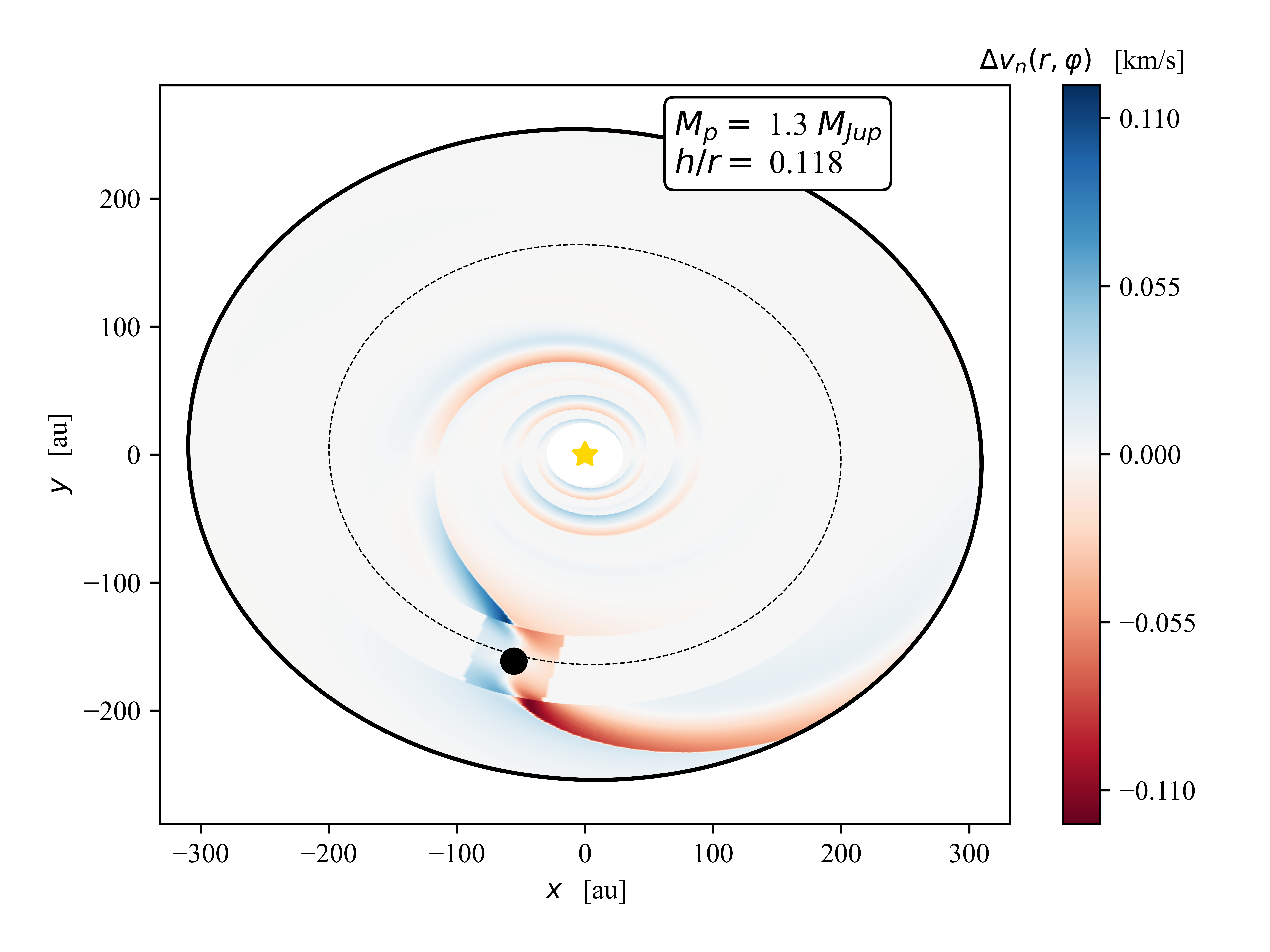}
    \includegraphics[width=8cm]{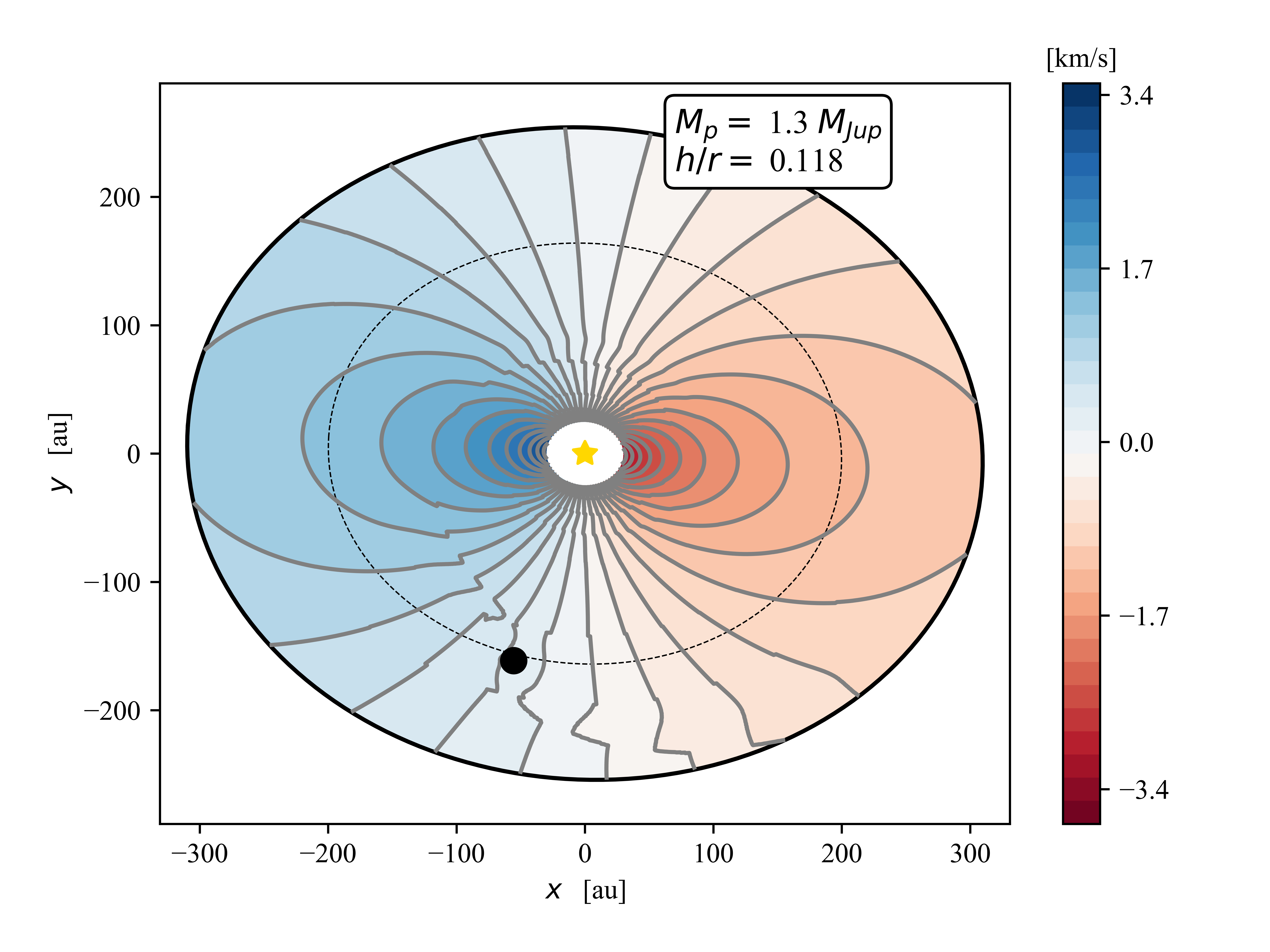}
    \caption{
    Output of the semi-analytic model for a planet mass of 1.3\,\mjup \ at 206\,au at a position angle of 170$^{\circ}$ and a disk scale height of 0.118 at the planet position (as in \citealt{Bae22}). The figure shows: the density perturbation (top), the deviation from keplerian velocity (middle) and the velocity perturbation (bottom) induced in the velocity channels map (channel width of 0.2\,\kms). The dashed line represents an orbital distance of 200\,au. 
    }
    \label{fig:bae}
\end{figure}

\subsection{Kink analysis}
The overall structure of $\Delta v$ resembles a kinematics perturbation (or kink) as those induced by a planet \citep[e.g.,][]{perez15,Pinte18,Pinte19, bollati21}. 
This is in good agreement with the recent results by \citet{Bae22}. We note however that the kink is not localized but it extends to almost the entire disk. \citet{Bae22} reports also the detection of a candidate gaseous circumplanetary disk in the $^{13}$CO channels map at $r=206\,$au and PA $\sim 170 - 190^{\circ}$. This structure is spatially coincident with the kink as shown in Figure~\ref{fig:diff} where the $^{13}$CO emission is overlaid on top of the 
$^{12}$CO one.
The kink clearly extends over the entire disk and in a number of channels it shows a characteristic wiggle in the radial direction.

\begin{table}[]
    \centering
    \begin{tabular}{llll}
    \hline\hline
    Parameter & Unit & Value & Description\\
    \hline
        $M_{*}$ & [$M_{\odot}$] & 1.2 & Stellar mass \\
        $d$ &[pc] & 121 & Heliocentric distance \\  
        $i$ & [$^{\circ}$] & 35 & Disk inclination \\
        PA &[$^{\circ}$] & 86 & Disk position angle\\
        $q$ && 0.25 & $c_s$ radial profile index \\
        $p$ && 1 & Index of surface density profile\\
        $\alpha m$ && 0 & Viscosity damping\\
        $\gamma$ && 5/3 & Adiabatic index \\
        $h/r$ && 0.11 & Disk scale height  \\
        \hline\hline
        \end{tabular}
    \caption{Input parameters to ``Analytical$\_$kinks$\_$master'' \citep[][]{bollati21}}
    \label{tab:models}
\end{table}

\section{Comparison to semi-analytical models}
The radial and azimuthal extent of the kink hints at the existence 
of a giant planet that perturbs the gas density distribution and kinematics. In this scenario, the kink originates at the intersection between the spiral wake and the line channels map.  
Beside disk-planet interaction, other mechanisms may also give rise to velocity perturbation, such as vertical shear instability (VSI) and gravitational instability (GI). However, the velocity perturbation produced by VSI is of the order of a few m/s \citep{Barraza21}, much smaller than the kink detected here. \citet{Longarini21} investigated the velocity perturbation induced by GI. The amplitude of the GI wiggle is consistent with the kink amplitude detected here but the global shape of the velocity field produced by GI differs substantially from the observed velocity map (Figure~\ref{fig:data_mom9}). 

In the following we assume the presence of a giant planet as a working hypothesis and we test whether this is capable of producing a wide-angle kink and, if so, to put constraints to the position and mass of such a planet. To do this we compare the observed kink to semi-analytic models by \citet{bollati21} based on the code Analytical$\_$kinks$\_$master\footnote{Available at https://github.com/fbollati/Analytical$\_$Kinks}.

\noindent
The radial and azimuthal extent of the perturbation in the semi-analytic model depends primarily on the planet mass (\pmass) and orbital distance (R$_{\rm P}$) and on the disk scale height 
h/r at $r =$ \prad ($h_{\rm P}$).
\citet{bollati21} included a viscosity parameter ($\alpha_m$) to artificially damps the spiral wake. As we observe a large scale kink, we imposed $\alpha_m = 0$. 
The $\alpha_m$ parameter is a free parameter that tunes the strength of the exponential viscous damping of velocity perturbations, irrespective of the nature of the viscosity. This parameter was first introduced in \cite{bollati21} to account for a possible impact of such dumping on the kinks observed in channel maps. However in the case of HD 163296 \cite{calcino22} showed that this velocity damping prescription is not required as the planet wake induces secondary kinks in velocity channels that extend far from the planet location, i.e. the semi-analytical model successfully reproduces the observations assuming $\alpha_m$=0. Being the velocity kinks in AS 209 not localized nearby the the planet but spread at all azimuths in a spiral-like shape, we decided to neglect viscous damping as well. 

In our model, the disk scale height is given by the ratio of the sound speed ($c_{\rm s}$) and the Keplerian velocity ($v_{\rm Kep}$). For the sound speed, we assume a power-law dependence with radius ($c_{\rm s} \sim r^{-q}$). Thus the scale height varies with the radius as $h/r \propto r^{0.5-q}$. We fix $q = 0.25$ (hence $h/r \sim r^{0.25}$) and the disk scale height (at $r = 100\,$au) to $h_{\rm P}$ = 0.118 \citep[e.g.,][]{law21}.

Once the viscosity damping and the scale height are fixed, the extent and the amplitude of density and velocity perturbation depend only on the planet position and mass.

\subsection{Planet position}
To estimate the planet position we run a grid of models  inspecting different values of \prad \ and planet position angle (\pap). We initially tested the parameters proposed by \citet{Bae22} (\pmass = 1.3\,\mjup, \prad = 206\,au, \pap~=170$^{\circ}$). The results are shown in Figure~\ref{fig:bae}. With these values the synthetic kink is localized in a few channels in the proximity of the planet and the model does not reproduce the large scale structure of the observed perturbation. In particular, the radial and azimuthal extent of the synthetic kink is not consistent with the observations. This is because the outer spiral arm extends to larger distances, beyond the disk radius.

\noindent
In a second step we tested different values of \pap \ while keeping fixed the orbital radius \prad = 200\,au. 
 In all these models, the extent of the kink does not reconcile with the observations. The only way to reproduce the tightly wound spiral of the observations is to impose a low scale height value at the planet position ($h_{\rm P} < 0.05$, Figure~\ref{fig:rafikov}). This is however in contradiction with previous estimates of the disk scale height based on the fit of the spectral energy distribution \citep[e.g.,][]{Andrews11} and of the CO emitting surfaces \citep[][]{Teague18b, law21}. Moreover, these simulations do not produce the wiggles observed at $r \sim 100-200\,$au.

\begin{figure}[!t]
    \centering
    \includegraphics[width=8cm]{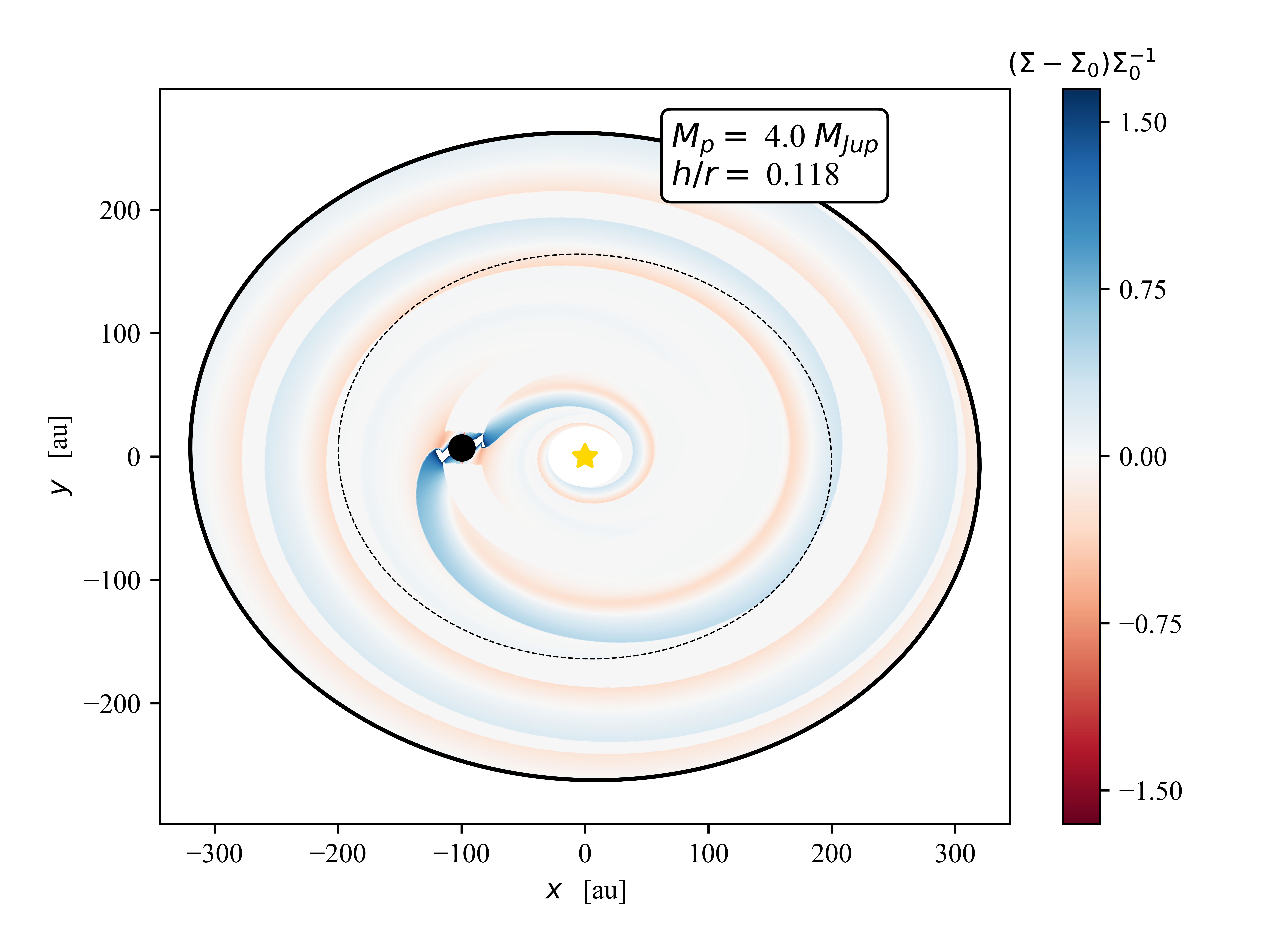}
    \includegraphics[width=8cm]{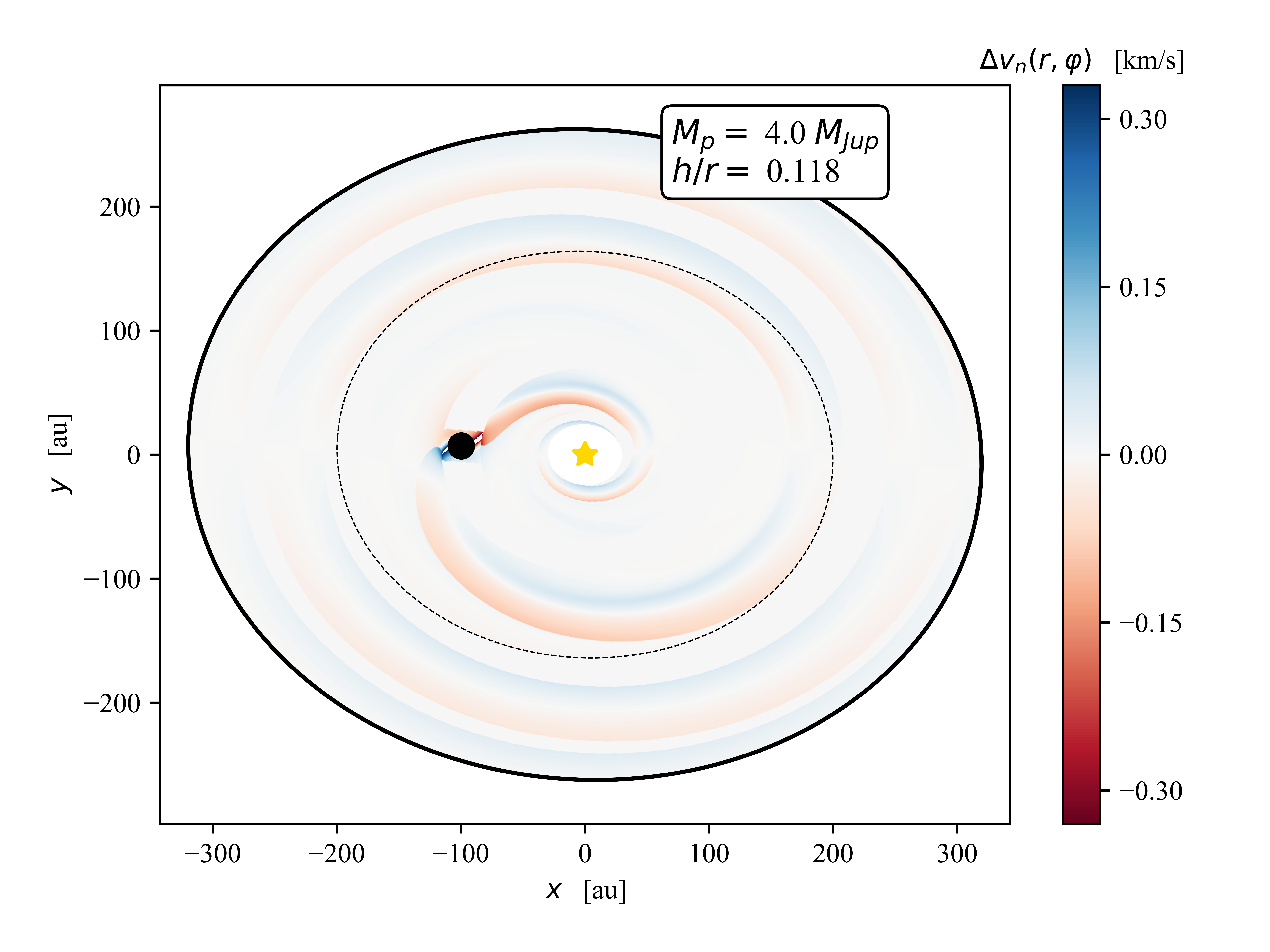}
    \includegraphics[width=8cm]{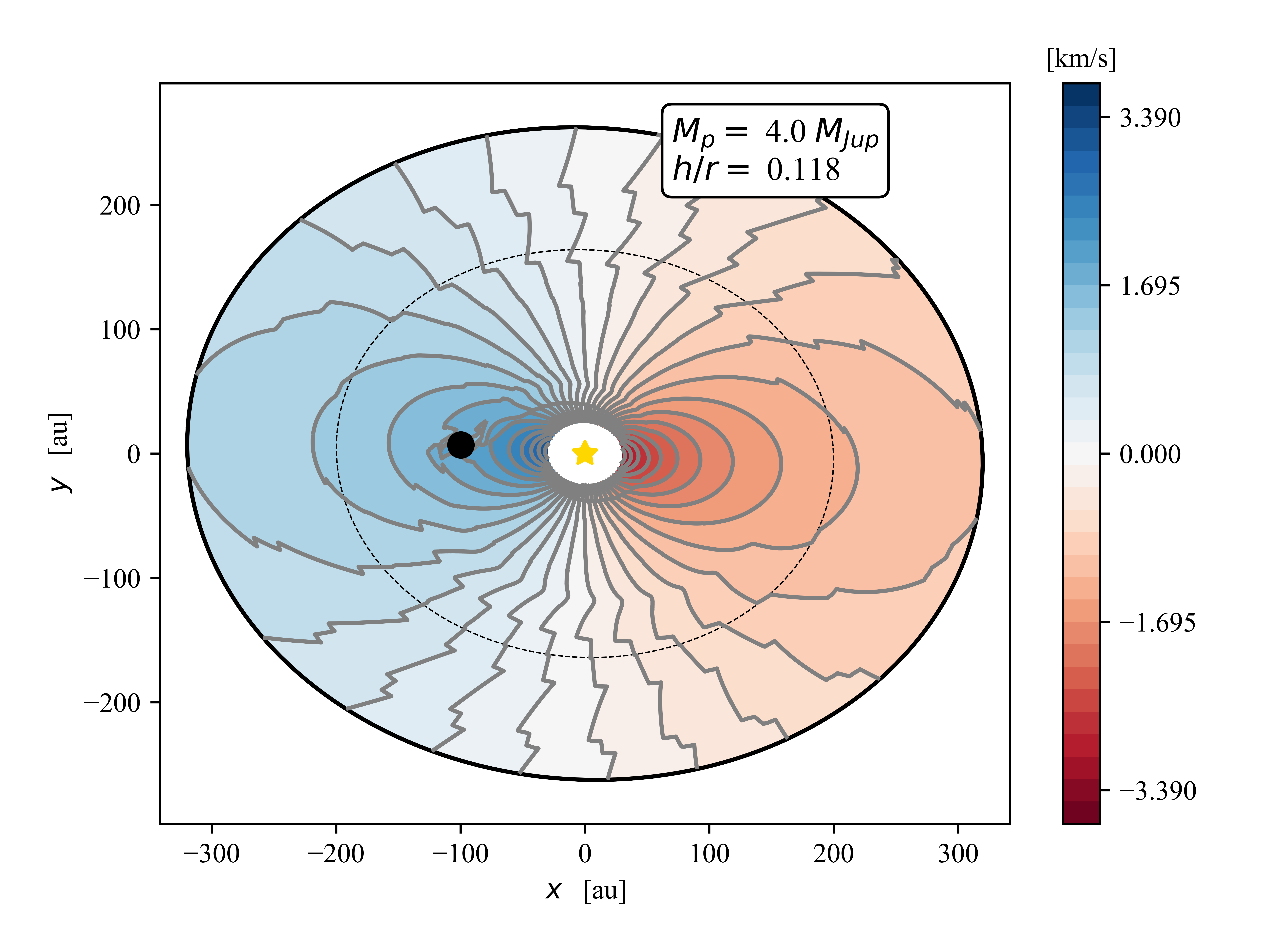}
    \caption{
    Same as Figure~\ref{fig:bae} for the representative model with \prad=100\,au, \pmass=4\,\mjup, \pap=90$^{\circ}$ and $h_{\rm P}$=0.118. }
    \label{fig:sim}
\end{figure}

We finally tested the case of \prad = 100\,au. The velocity perturbations in Figure~\ref{fig:data_mom9} are very extended both radially and azimuthally and multiple wiggles are clearly identified in each given channel. This may suggest that the outer spiral arm folds more than one time, and for this to happen the planet must be located further in. The presence of a planet at $\sim$ 100\,au has been proposed to explain the gap at observed in the dust \citep[e.g.,][]{Fedele18, andrews18a} and gas distribution \citep[][]{favre19}. As a test, Figure~\ref{fig:rafikov} shows the analytic wake \citep[][]{Rafikov02} for \prad~= 100\,au, \pap~=90$^{\circ}$ and $h_{\rm P}$ = 0.118 overlaid on the $^{12}$CO velocity map. Notably, the outer spiral arm folds multiple times crossing the velocity channels close to the the observed kinks. With the assumption of \prad~=100\,au, we run a new grid of models by varying \pmass \ and \pap . Models with \pap \ $\sim 60^{\circ} - 110^{\circ}$ (East of North) produce a  pattern similar to the observed one. As an example, Figure~\ref{fig:sim} shows the predictions of the semi-analytic model for the density and velocity perturbation and the synthetic velocity map for the case of \pmass~= 4\,\mjup \ and \pap~=90$^{\circ}$. The outer spiral arm crosses the velocity channels multiple times, giving rise to several wiggles in each given channel. The radial extent and amplitude (azimuthal offset) of the kink produced by the outer arm increases as the wake moves away from the planet. In the proximity of the planet, the perturbation is smaller than the channel width of 0.2\,\kms. All these findings are in agreement with the observed velocity perturbation strengthening the hypothesis of  \prad~=100\,au. In this scenario, the previously detected gap at $r \sim 200\,$au is most likely due to density fluctuations induced by the outer spiral arm (Figure~\ref{fig:sim}, top panel).  

\begin{figure}[!t]
    \centering
    \includegraphics[width=5cm]{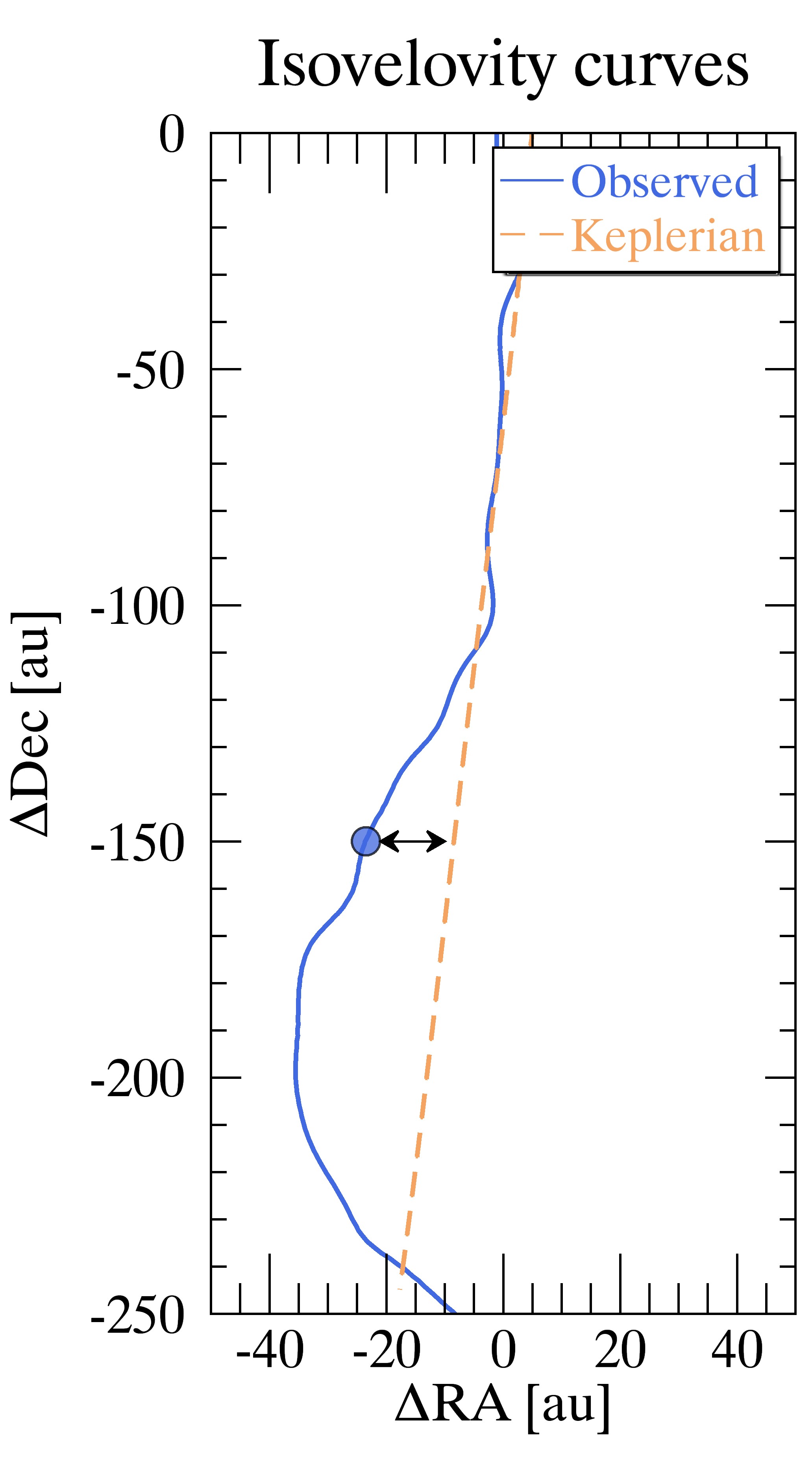}
    \caption{Observed vs Keplerian isovelocity curves in the central channel. The central channel is adopted as reference to measure the kink amplitude ($\mathcal{A}$) as the offset between the observed and keplerian velocity. The Keplerian profile is based on eq.~\ref{eq:l21} for the top surface layer. The two dots indicate the position where the offset is measured.}
    \label{fig:offset}
\end{figure}

\subsection{Planet mass}
\begin{figure}[!t]
    \centering
    \includegraphics[scale=0.5]{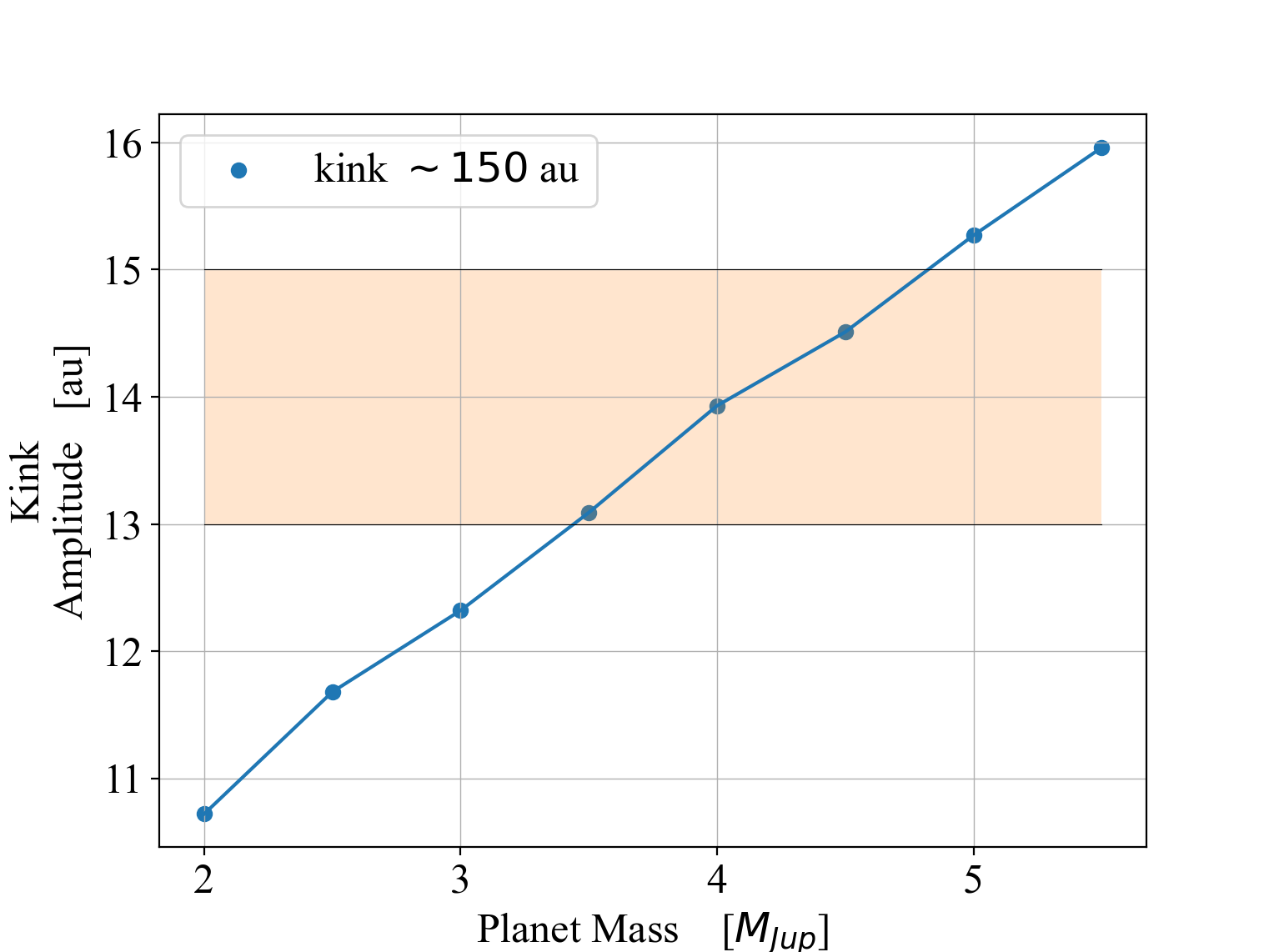}
    \caption{Model prediction of the kink amplitude as a function of the planet's mass, assuming that the planet is located at $r=100\,$au and PA = 90$^{\circ}$ and with a disk scale height of $h/r = 0.118$ at R$_{\rm P}$. Masses that best reproduce the observe kink amplitude fall in the range $\sim 3.5 M_\textrm{Jup} - 4.5  M_\textrm{Jup}$.}
    \label{fig:sim_offset}
\end{figure}

To estimate \pmass \ we created a new simulation grid with fixed \prad \ (100\,au), \pap \ (90$^{\circ}$) and $h_{\rm P}$ (0.118). In each given channel the amplitude of the kink ($\mathcal{A}$ [au]) increases with \pmass.  We measured $\mathcal{A}$ as the offset between the Keplerian and the observed central velocity and we compared it to the offset in the synthetic velocity map. Figure~\ref{fig:offset} shows the observed ($v_{\rm o}$) and the Keplerian ($v_{\rm k}$) isovelocity curves of the central channel of $^{12}$CO. The Keplerian profile is based on eq.~\ref{eq:l21} and it refers to the top CO surface layer as fitted by \citet{law21}. We inspected the central channel at $v = -4.7\,$\kms (corresponding to the disk minor axis) as to avoid projection effect of the velocity in the azimuthal direction. We measured $\mathcal{A}$ as a function of radial projected distance from the star and for the comparison with the synthetic map, we considered the offset at a projected distance of $r = 150\,$au which corresponds to the first intersection of the outer spiral arm with the central velocity channel. We note that at larger radial distances, the measurement of the observed offset can be overestimated because of the convolution of 2 consecutive kinks, as suggested by the semi-analytic simulations (Figure~\ref{fig:sim}, middle panel). The observed offset at $r = 150\,$au is $\mathcal{A}$ $\sim$ 13-15\,au . Figure~\ref{fig:sim_offset} shows the model prediction of the kink amplitude as a function of the planet's mass from the simulation grid. Based on this analysis we estimate a mass of \pmass $\sim$ 3.5 -- 5\,\mjup.

\subsection{Caveats}\label{sec:caveats}
The estimate of the planet mass and position angle based on the semi-analytical model by \citet{bollati21} has some limitations. The kink amplitude at the disk minor axis depends on the actual position angle of the planet and where we actually measure the offset. 
We also note that the planet-induced perturbation are computed at the midplane of the disk ($z=0$) and not at the actual height of the CO emitting surface. However, because of the vertical temperature gradient, the spiral can propagate differently at different heights as demonstrated by \citet{Juhasz18} and \citet{Rosotti20}. The pitch angle of the spiral is expected to increase from the midplane to the higher disk layers because of the higher gas temperature. Modelling the vertical temperature gradient is beyond the scope of this paper. For a detailed discussion on the limitation of the the models used here see \citet{calcino22}.

\noindent
The remaining parameters of the semi-analytical model ($q, p, \gamma$) have a minor effect on the estimate of the planet mass \citep[see][]{bollati21}.

\section{Conclusions}

In this paper, we have presented a clear evidence of a kink in the velocity field of the disk in AS 209. 
We confirm the presence of the kink previously detected by \citet{Bae22} in the $^{12}$CO $J=2-1$ velocity map and we report the detection of the kink in the $^{13}$CO $J=2-1$ map as well. The kink is detected throughout the entire disk at different orbital distances from the star. In the southern region, multiple wiggles are detected with a coherent structure in consecutive channels, consistent with a spiral wake which folds multiple times. 
A deep gap (both in the dust and gas distribution) and a velocity kink are two of the most evident indication of planet-disk interaction and the simultaneous occurrence of both strongly points towards the presence of an embedded planet at $\sim$ 100\,au.  
We have compared the observed amplitude of the kink to the expectations based on analytical models of velocity perturbations due to planet-disk interaction and we conclude that the putative planet orbits at 100\,au from the central star and it has a mass between $\sim$ 3 and 5\,\mjup. Our estimate of \pmass \ exceeds by an order of magnitude the value reported in \citet{Fedele18}, \citet{zhang18} and \citet{favre19} who compared the gap properties to hydrodynamic simulations. We note however, that while the gap width and depth in hydrodynamic simulations do not provide a unique solution for \pmass , in the case of velocity perturbations, a lower mass planet cannot produce the observed kink amplitude.    

The presence of a massive planet at 100\,au poses several challenging questions. 
Unimpeded radial drift is known to lead to very compact dust disks, much more compact than observed. 
 
The common explanation for this discrepancy is to invoke disk substructures as a way to slow down dust drift and to help the formation of planetesimals followed by pebble accretion \citep[e.g.,][]{Lambrechts14}. As an example, \citet{Bae15} proposed the formation of vortices during the infall of the protostellar envelope. 
Another possible solution is the formation of narrow dust rings at several tens of au by magnetically coupled disk winds \citep{Suriano18}. 

As an alternative, a planet at such a large distance can form by gravitational instability, either by rapid solid core formation induced by the instability \citep{Rice04}, or by direct gas fragmentation \citep{Rice05}. In the case of disk fragmentation, the theoretical predictions of the initial fragment mass at 100\,au goes from $\sim 2-20\,$\mjup \ \citep[e.g.,][and references therin]{kratter16}, in good agreement with our estimate.

The evidence gathered so far places AS 209 as a prime candidate for direct detection of the putative planet in the infrared by either ground or space based telescopes, such as VLT/ERIS and JWST.

\begin{acknowledgements}
We thank the anonymous referee for the thoughtful report and comments that helped to improve the paper. 
DF acknowledges the support of the Italian National Institute of Astrophysics (INAF) through the INAF Mainstream projects ARIEL and the ``Astrochemical Link between Circumstellar Disks and Planets", ``Protoplanetary Disks Seen through the Eyes of New- generation Instruments" and by the PRIN-INAF 2019 Planetary Systems At Early Ages (PLATEA). This project has received funding from the European Union’s Horizon 2020 research and innovation programme under the Marie Skłodowska-Curie grant agreement No 823823 (DUSTBUSTERS).

\end{acknowledgements}


\begin{appendix}
\section{JvM uncorrected data}
\begin{figure}
    \centering
    \includegraphics[width=9cm]{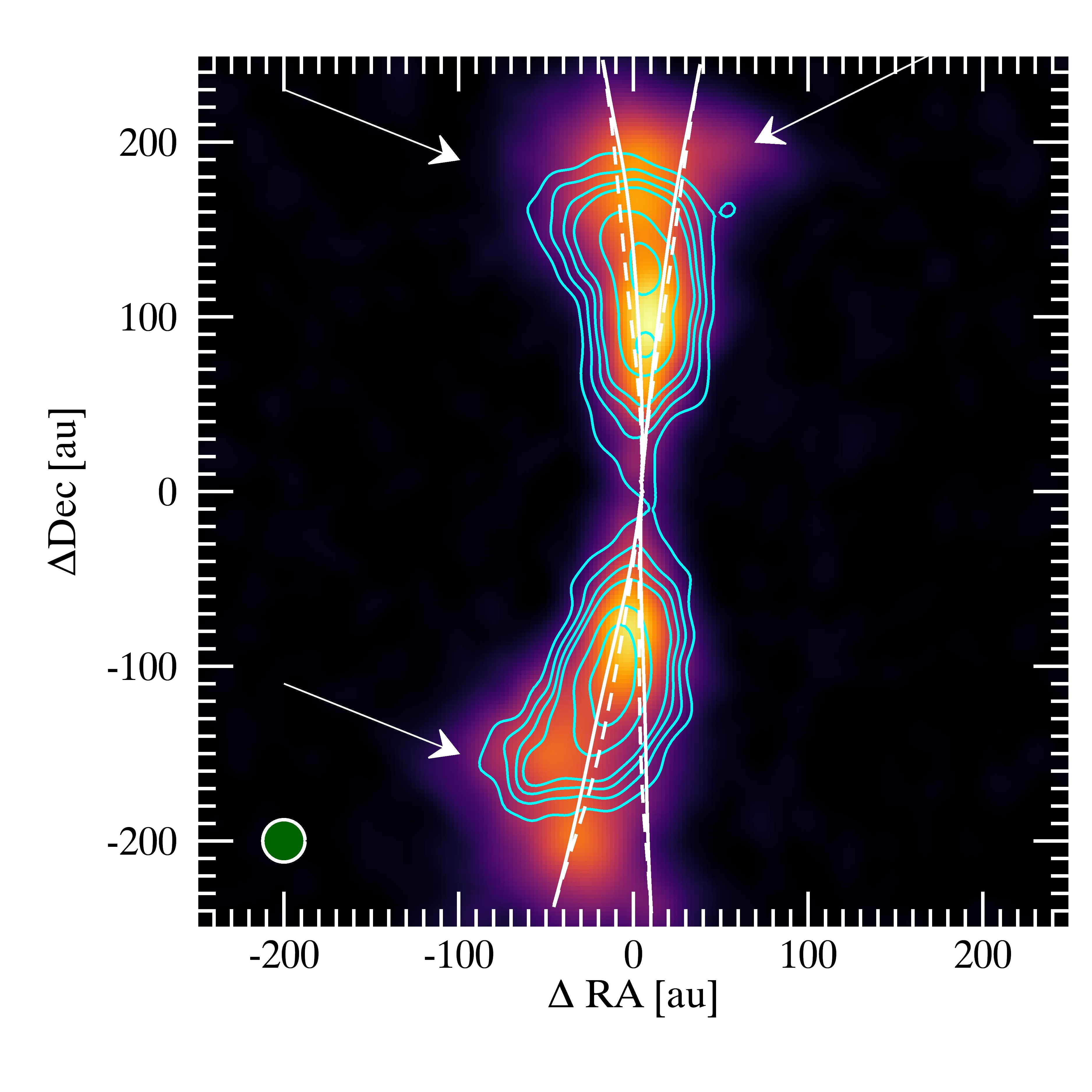}
    \includegraphics[width=9cm]{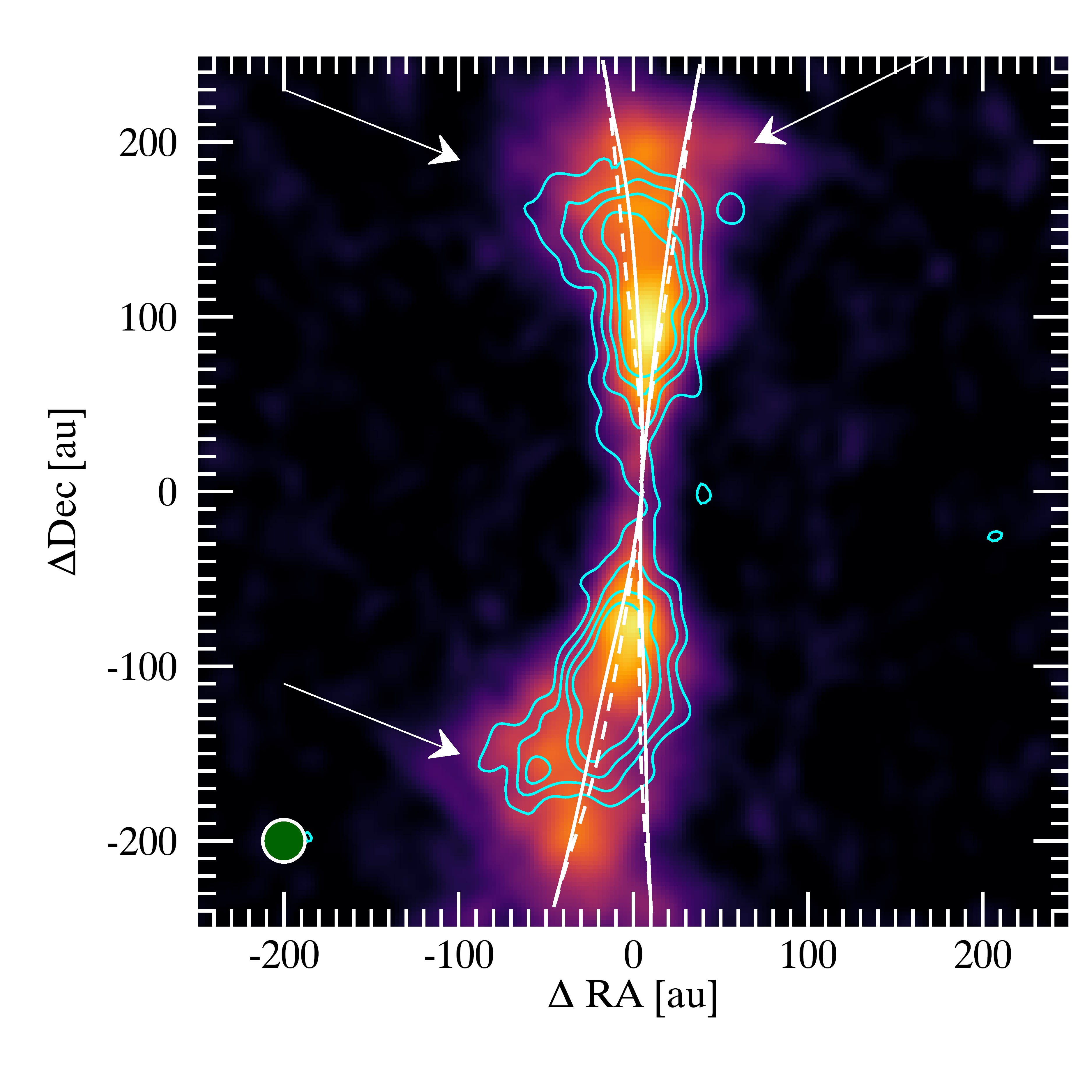}
    \caption{JvM-corrected (top) and uncorrected (bottom) image of the central velocity channel of $^{12}$CO and $^{13}$CO (cyan contours). The (white) lines represent the expected contours in the case of Keplerian profile for the top (solid) and bottom (dashed) surface (eq.~\ref{eq:l21}). The arrows point to the main deviation from Keplerian velocity. The restored beam of $0\farcs2$ is shown in the bottom left corner as a green circle. The $^{13}$CO contours are plotted at 3, 5, 7, 9, 15 and 20\,$\sigma$.}
    \label{fig:diff}
\end{figure}

\citet{Casassus22} warned that the JvM correction \citet{Jorsater95} can exaggerate the peak signal-to-noise of the restored image. To ensure the reliability of our results we analysed the MAPS cleaned image without the JvM correction. Figure~\ref{fig:diff} shows a comparison between the JvM-corrected (top) and JvM-uncorretted (bottom) images of the central channel. The $^{13}$CO contours are overlaid ontop of the $^{12}$CO image for a direct comparison. Note that for this comparison we used the images restored with a beam size of $0\farcs$2, slightly higher than the data used in the analysis above. 
The peak signal-to-noise is indeed higher in the first case. Nevertheless, the kink is clearly visible also in the images without the JvM correction. We note in particular that the kink is detected in multiple positions (in both lines) also in the JvM uncorrected dataset.

\section{Analytic spiral wake in linear regime}
\begin{figure}
    \centering
    \includegraphics[width=9cm]{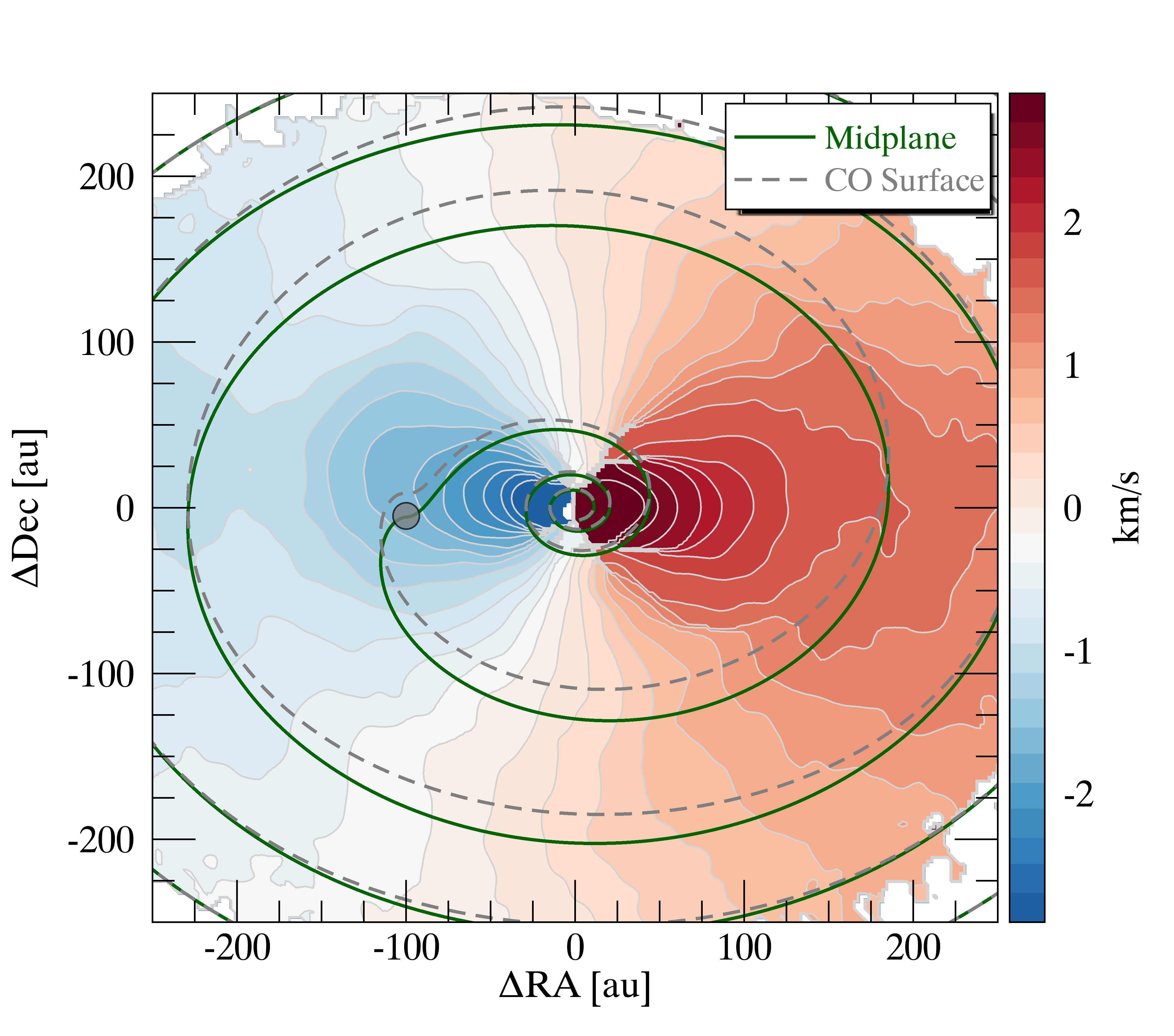}
    \includegraphics[width=9cm]{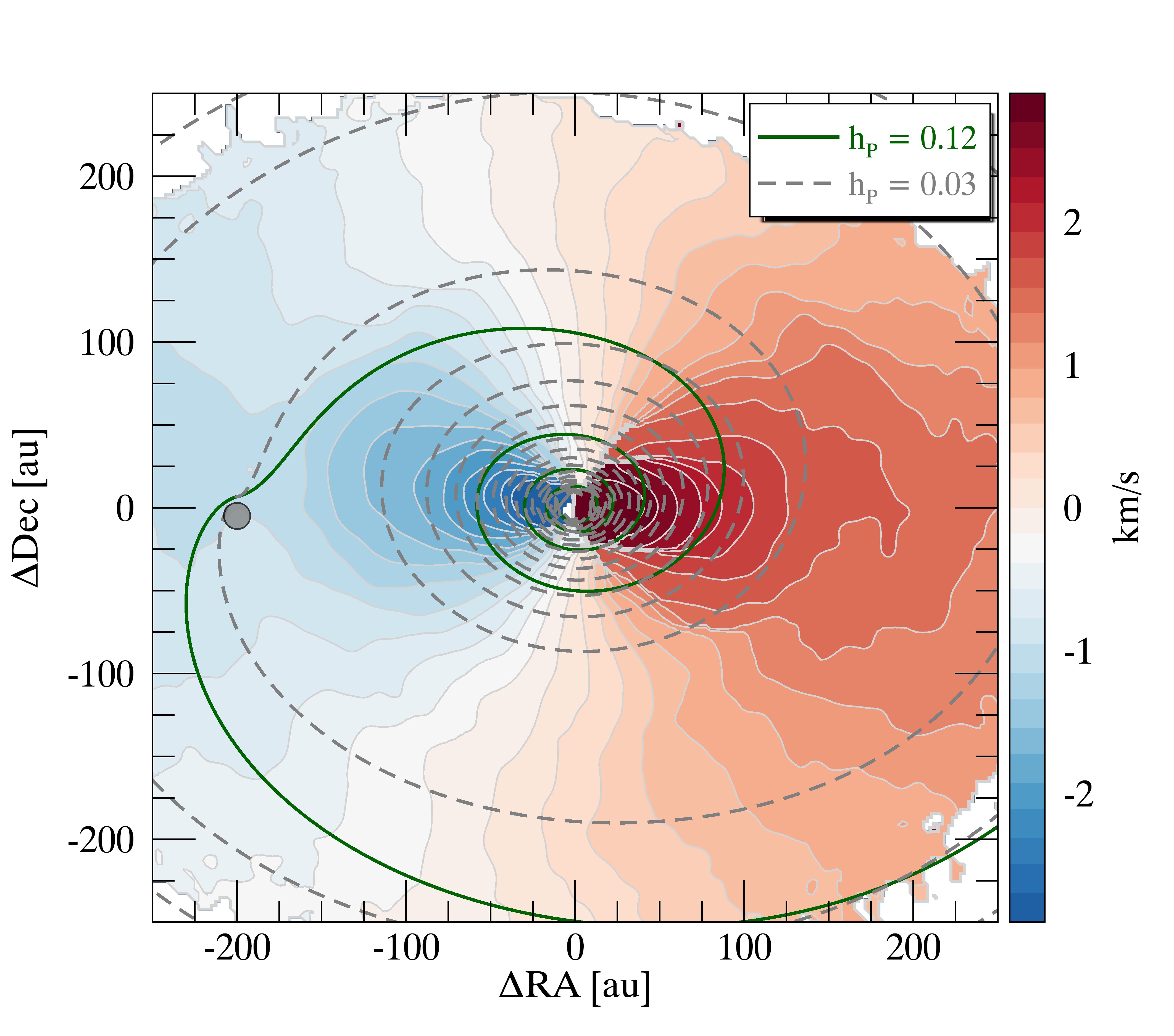}
    \caption{Analytic spiral wake overlaid on top of the $^{12}$CO velocity map. The grey dot indicate the position of the planet at the midplane. ({\it top}) The planet orbital radius is 100\,au and the scale height is $h_{\rm P}=0.12$. The two curves show the spiral wake at the midplane (green solid) and the CO top surface layer (dashed grey). ({\it bottom}) The planet is at \prad=200\,au and the spiral (projected on the CO top surface layer) is shown for two different values of $h_{\rm P}$.}
    \label{fig:rafikov}
\end{figure}

Figure~\ref{fig:rafikov} shows the analytic spiral wake on top of the observed $^{12}$CO velocity map. The analytic spiral wake is computed the spiral wave in the linear regime following \citet{Rafikov02}. The planet is at $r=100\,$au and PA=90$^{\circ}$, the disk scale height is $h_{\rm P}$=0.118, the power exponent of the radial gradient of the sound speed velocity is 0.35. The spiral is computed at the midplane (dashed black line) and than projected on the CO top surface layer as defined by eq.~\ref{eq:l21}, respectively. In several channels, the projected spiral wake matches the position of observed kinks. The projected spiral does not take into account the vertical temperature gradient (see sec.~\ref{sec:caveats}).

\end{appendix}

\end{document}